\documentclass[aps,english,10pt,superscriptaddress,twocolumn]{revtex4}
\pdfoutput=1
\usepackage{amsmath}
\usepackage{amsthm}
\usepackage{amssymb}
\usepackage{amsfonts}
\usepackage{bbm}
\usepackage{epsfig}
\usepackage{times}
\usepackage{babel}
\usepackage{color}
\usepackage{hyperref}
\usepackage{framed}
\usepackage{changes}
\usepackage{float}
\usepackage{array}
\usepackage{tikz}
\usetikzlibrary{backgrounds}
\usepackage{comment}
\usepackage[T1]{fontenc} 
\usepackage[textsize=footnotesize,textwidth=2.5cm]{todonotes}

\def \be {\begin{equation}}
\def \ee {\end{equation}}
\def \ba {\begin{array}}
\def \ea {\end{array}}
\def \bea{\begin{eqnarray}}
\def \eea{\end{eqnarray}}

\hypersetup{
colorlinks=false,
linkcolor=black
}

\makeatother

\begin{document}

\title{Surface/State correspondence and $T\overline{T}$
deformation}

\author{Bin Chen}
\email{bchen01@pku.edu.cn}
\affiliation{Department of Physics and State Key Laboratory of Nuclear
Physics and Technology, Peking University, No.5 Yiheyuan Rd, Beijing
100871, P.R. China}
\affiliation{Center for High Energy Physics, Peking University,
No.5 Yiheyuan Rd, Beijing 100871, P. R. China}
\affiliation{Collaborative Innovation Center of Quantum Matter,
No.5 Yiheyuan Rd, Beijing 100871, P. R. China}

\author{Lin Chen}
\email{linchen91@pku.edu.cn}
\affiliation{Department of Physics and State Key Laboratory of Nuclear
Physics and Technology, Peking University, No.5 Yiheyuan Rd, Beijing
100871, P.R. China}

\author{Cheng-Yong Zhang}
\email{zhangcy@jnu.edu.cn}
\affiliation{Department of Physics and Siyuan Laboratory, Jinan University, Guangzhou 510632, China}


\begin{abstract}
{The surface/state correspondence suggests that the bulk co-dimensional two surface could be dual to the quantum state in the holographic conformal field theory(CFT).
 \textcolor{black}{Inspired by the cutoff-AdS/$T\overline{T}$-deformed-CFT correspondence,} we propose that the quantum states of two-dimensional $T\overline{T}$-deformed holographic CFT {are} dual to some particular surfaces in {the AdS$_3$ gravity}. In particular, {the time slice of} the cut-off surface is dual to the ground state of the $T\overline{T}$-deformed CFT. We examine our proposal by studying the entanglement entropy and quantum information metric. We find that  the complexity of the ground state in the  deformed theory is consistent with the one of {a particular} cMERA and the holographic complexity via CV or CA prescription.}

\end{abstract}

\maketitle

\section{Introduction}

The study of the black hole thermodynamics inspired G. 't Hooft and  L. Susskind \cite{Hooft1993,Susskind1994} to propose holography
 as the guiding principle of quantum gravity.
One  concrete realization
of the holographic principle is the AdS/CFT correspondence, which states that
quantum gravity
theory in anti-de Sitter (AdS) spacetime could be dual to  a conformal field theory (CFT) living on the asymptotical  AdS boundary  \cite{Maldacena1998}.
Among various issues in AdS/CFT, the emergence of bulk geometry  is outstanding.
One promising proposal is the
 surface/state correspondence
\cite{Miyaji2015b,Miyaji2015a}, which claims that any co-dimension
two convex surface $\Sigma$ corresponds to a quantum state described
by a density matrix $\rho(\Sigma)$ \textcolor{black}{in the dual Hilbert space}.
The proposal is
based on the recent development of holographic entanglement entropy \cite{Ryu2006a,Ryu2006b} and interesting connection between the
AdS/CFT and the tensor network \cite{Swingle2009,Nozaki2012}. The physical picture of the surface/state
correspondence is intuitive and has inspired many new \textcolor{black}{discoveries}, such
as the holographic entanglement of purification \cite{Takayanagi2017}.
However, the concrete form of the correspondence  is still
far from clear.

In this paper, we study the surface/state correspondence in the context of
 newly discovered cutoff-AdS/$T\overline{T}$-deformed-CFT (cAdS/dCFT)
correspondence \cite{McGough2016}.  The study of $T\overline{T}$ deformation was pioneered by
Smirnov and Zamolodchikov \cite{Zamolodchikov2004,Smirnov2016}. They
studied \textcolor{black}{the deformation triggered
by the composite operator} $T_{zz}T_{\bar{z}\bar{z}}-T_{z\bar{z}}^{2}$ in two-dimensional (2D) quantum field theory,
and found that an integrable quantum filed theory  is still integrable after the $T\overline{T}$
deformation. More interestingly, it was conjectured that
the $T\overline{T}$-deformed holographic CFT$_2$ in the large central charge limit could be dual to the cutoff AdS$_3$ gravity provided
that \textcolor{black}{the sign of} the deformation parameter is chosen suitably \cite{McGough2016}. For some related works,
see  \cite{Giribet:2017imm,Kraus2018,Bonelli2018,Caputa2019,Hartman2018}.
As the boundary moves into the bulk, one gets a series of codimension-one hypersurfaces. These hypersurfaces are exactly where the $T\overline{T}$-deformed CFTs live. Their codimension-two time slices are where the
quantum states of the $T\overline{T}$-deformed CFTs \textcolor{black}{dwell}. This fact inspires us to propose that the surface/state correspondence can be realized
by the cAdS/dCFT correspondence.
{These co-dimension two time-slices in AdS$_3$ correspond to the vacuum
states of the $T\overline{T}$-deformed CFTs, while those in the BTZ background correspond to the mixed states at finite temperature.} We will examine this proposal quantitatively
by studying the entanglement entropy and complexity in both field theory and bulk gravity.

Entanglement entropy plays an important role in modern physics, especially
in quantum information theory and quantum gravity. The milestone development
is the holographic entanglement entropy.
The authors  in \cite{Ryu2006a,Ryu2006b} proposed that the entanglement entropy in holographic CFT can be calculated
holographically by the area of the bulk minimum surface anchoring on the boundary entanglement surface.
The appearance of minimal surface, which is called the RT-surface,  encouraged people to
conjecture that the spacetime may \textcolor{black}{emerge} from the quantum entanglement
\cite{Raamsdonk2010}. Another piece of evidence supporting this the conjecture comes from the
analog between tensor network \cite{Vidal2005} and the RT surface \cite{Swingle2009}.  The connection between tensor network and holographic entanglement surface led to the works on the spacetime reconstruction from tensor network \cite{Swingle2012},
which further led to the surface/state correspondence \cite{Miyaji2015b,Miyaji2015a}.
Thus it is very interesting and necessary to examine our proposal
from the viewpoint of entanglement entropy.

Though entanglement plays a crucial role in studying the dual gravity,
it is not enough \cite{Susskind2014}, especially when one tries to
understand the interior of the black hole. The complexity of
a quantum state was thus introduced as a dual probe to investigate the growth
rate of the Einstein-Rosen bridge \cite{Susskind2014,Susskind2014a}.
There have been two conjectures about the holographic complexity: the CV conjecture \cite{Stanford2014}
and CA conjecture \cite{Brown2015}.
{These two conjectures share some similarities but differ in many ways \cite{Carmi2017,Yang2017}. More importantly, how to define a
 complexity in quantum field theory is still under intense investigations \cite{Nielsen2006,Chapman2017,Jefferson2017,Caputa2017a,Caputa2017b,Molina2018,Yang2018a,Yang2018b,Yang2019}. }
Since we are studying the surface/state correspondence which was inspired
from the tensor network, we will adopt the definition from \cite{Chapman2017,Molina2018}.

Our study could be seen from another point of view.
In \cite{Miyaji2015c},  the quantum information metric in a CFT with respect to a small marginal deformation has been studied. Here we study the quantum information metric with respect to an irrelevant deformation.


\section{Surface/State correspondence, $T\overline{T}$ deformation and entanglement}


\subsection{Surface/State correspondence}

\begin{figure}[htbp!]
\centering{}\includegraphics[width=0.16\textwidth]{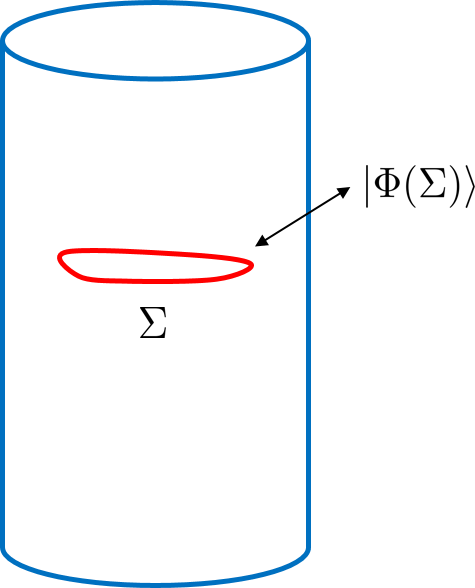} \caption{\label{ssdual} A codimension two surface
$\Sigma$ corresponds to a quantum state $|\Phi(\Sigma)\rangle$.}
\end{figure}

Let's briefly review the surface/state correspondence \cite{Miyaji2015b,Miyaji2015a}.
It claims that
any codimension two convex surface $\Sigma$ corresponds to a quantum
state described by a density matrix $\rho(\Sigma)$ in the dual Hilbert
space. When this surface is closed and topologically trivial, the
state is given by a pure state $|\Phi(\Sigma)\rangle$, as shown in
Figure \ref{ssdual}. Especially, if $\Sigma$ is a time silce of
AdS boundary, $|\Phi(\Sigma)\rangle$ is simply the ground state of
the dual CFT.

\begin{figure}[htbp!]
\centering{}\includegraphics[width=0.14\textwidth]{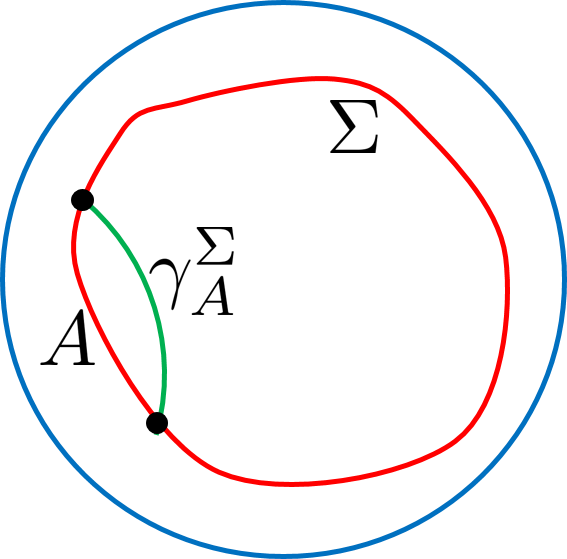} \caption{\label{grtformula} Generalized Ryu-Takayanagi formula.}
\end{figure}

A subregion $A$ of $\Sigma$ has its own dual state $\rho_{A}$,
and we can define the entanglement entropy of $A$ as
$
S_{A}^{\Sigma}=-\text{tr}_{A}\rho_{A}\log\rho_{A}.
$
It was proposed  \cite{Miyaji2015b,Miyaji2015a} that $S_{A}^{\Sigma}$ satisfies the generalized Ryu-Takayanagi
formula:
$
S_{A}^{\Sigma}=\frac{\textrm{Area}(\gamma_{A}^{\Sigma})}{4G_{N}},
$
where $\gamma_{A}^{\Sigma}$ is a minimal surface in the bulk whose
boundary is given by $\partial A$, as shown in Figure \ref{grtformula}.

\subsection{$T\overline{T}$-deformed CFT}

\begin{figure}[htbp!]
\centering{}\includegraphics[width=0.25\textwidth]{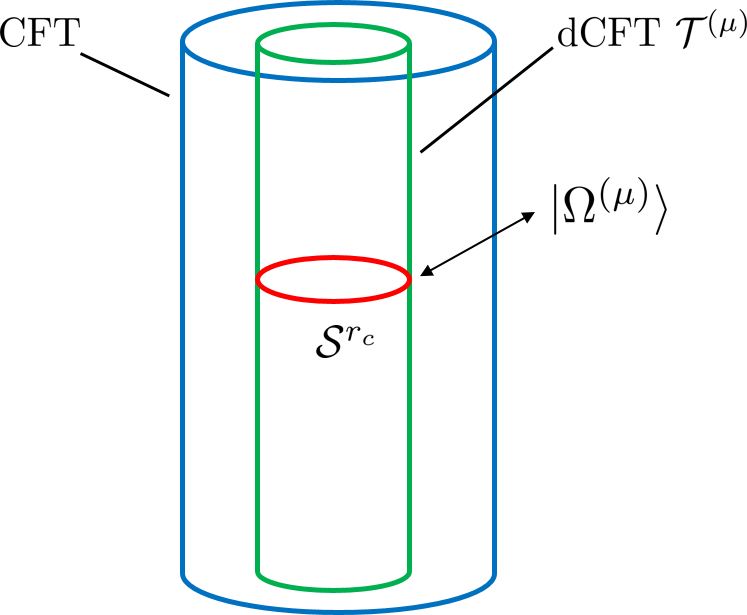}\caption{\label{dcft} Quantum state $|\Omega^{(\mu)}\rangle$ of $\mathcal{T}^{(\mu)}$
lives on $\mathcal{S}^{r_{c}}$.}
\end{figure}

When a 2-dimensional quantum field theory is deformed by the $T\overline{T}$
\textcolor{black}{operator}, we obtain a family of theories $\mathcal{T}^{(\mu)}$.
\textcolor{black}{Moving infinitesimally} from $\mathcal{T}^{(\mu)}$ to $\mathcal{T}^{(\mu+\delta\mu)}$, the action \textcolor{black}{changes} as
\begin{align}
S^{(\mu)} & \to S^{(\mu+\delta\mu)}=S^{(\mu)}+\delta\mu\int d^{2}xO^{(\mu)},\\
O^{(\mu)} & \equiv T^{(\mu)}\bar{T}^{(\mu)}-\Theta^{(\mu)2},
\end{align}
 where
$
T^{(\mu)}=-2\pi T_{zz}^{(\mu)},\  \bar{T}^{(\mu)}=-2\pi T_{\bar{z}\bar{z}}^{(\mu)},\ \Theta^{(\mu)}=2\pi T_{z\bar{z}}^{(\mu)}
$
 are the energy-momentum tensor of theory $\mathcal{T}^{(\mu)}$. The
deformation parameter $\mu$ has dimension $(\mbox{Length})^{2}$ and \textcolor{black}{this deformation} is irrelevant. The original theory \textcolor{black}{sits at} $\mu=0$. If the quantum
field theory is \textcolor{black}{a} holographic CFT, the cAdS/dCFT correspondence claims that \textcolor{black}{the deformed theory} $\mathcal{T}^{(\mu)}$
is dual to a gravitational theory living in a finite region in AdS$_{3}$ with a radial
cutoff at $r=r_{c}$ where the metric can be written as
\begin{equation}\label{AdSmetric}
ds^{2}=\frac{L^{2}}{r^{2}}dr^{2}+\frac{r^{2}}{L^{2}}g_{\alpha\beta}dx^{\alpha}dx^{\beta},\;\;(r\leq r_{c}).
\end{equation}
 Here $L$ is the AdS radius.  Within our convention, $\mu$
and $r_{c}$ are related by
\begin{equation}
\mu=\frac{6L^{4}}{\pi cr_{c}^{2}}.\label{muandrc}
\end{equation}
The theory $\mathcal{T}^{(\mu)}$ lives on the new boundary at $r=r_{c}$
which is a codimension one surface. The time slice $\mathcal{S}^{r_{c}}$
of the new boundary is a codimension two surface on which the quantum
states of $\mathcal{T}^{(\mu)}$ \textcolor{black}{dwell}, as shown in Figure \ref{dcft}.

\subsection{Entanglement entropy}

The similarity between Figure \ref{ssdual} and Figure \ref{dcft}
inspires our proposal: cAdS/dCFT is one kind of realization of the
surface/state correspondence. The state corresponding to \textcolor{black}{the} surface $\mathcal{S}^{r_{c}}$
is the quantum state of theory $\mathcal{T}^{(\mu)}$. Especially,
when the $T\overline{T}$ deformation is performed on the CFT ground
state, the surface $\mathcal{S}^{r_{c}}$ will correspond to the ground
state $|\Omega^{(\mu)}\rangle$ of $\mathcal{T}^{(\mu)}$, which could
be expressed as
\begin{equation}
\langle\varphi|\Omega^{(\mu)}\rangle=\frac{\int_{\phi(t=0)=\varphi}D\phi\;e^{-\int_{-\infty}^{0}dt\int dx\mathcal{L}^{(\mu)}}}{(\int D\phi\;e^{-\int dt\int dx\mathcal{L}^{(\mu)}})^{\frac{1}{2}}}.
\end{equation}
 Here $\mathcal{L}^{(\mu)}$ is the Euclidean Lagrangian density
of theory $\mathcal{T}^{(\mu)}$, $t$ and $x$ are the Euclidean coordinates.
The boundary condition should satisfy $\phi(t=0)=\varphi$. Note that
through $T\overline{T}$ deformation, we can only \textcolor{black}{obtain} the corresponding
states of surfaces like $\mathcal{S}^{r_{c}}$ which \textcolor{black}{comes} from the
uniformly moving inward of the original boundary. The corresponding
states of more general codimension two \textcolor{black}{surfaces} are \textcolor{black}{still} unknown.

In our study we do not need to require the existence of  a BTZ black hole in the bulk. This means that we focus on the ground state of $T\overline{T}$-deformed holographic CFT at zero temperature. Without the black hole, the metric \eqref{AdSmetric} is invariant under the scaling  $r\to \lambda r,t \to t/\lambda,x\to x/\lambda$, and the cutoff theory is invariant under the rescaling $\ensuremath{r_{c}\to r_{c}/\lambda,t\to\lambda t,x\to\lambda x}$.
Since $\mu\propto r_{c}^{-2}$, we see that $(\mu,t,x)$ is equivalent
to $(\lambda^{2}\mu,\lambda t,\lambda x)$.
We may certainly include an BTZ black hole in the bulk. In this case the time-slice $\mathcal{S}^{r_{c}}$ corresponds to a mixed state {describing} the $T\overline{T}$-deformed CFT at a finite temperature.

In \cite{Chen2018} it was shown  that the RT formula still
\textcolor{black}{holds} for cAdS/dCFT  to the leading order of the deformation parameter.
Subsequently the study has been generalized to multiple intervals \cite{Jeong2019}
at finite temperature. More works on the entanglement entropy showed
that the RT formula still works in cAdS/dCFT \cite{Donnelly2018,Park2018,Chakraborty2018,Banerjee2019,Murdia2019,Ota2019}.
{Now the holographic entanglement entropy is captured by the length of the geodesic ending on the cut-off surface.
This is exactly the generalized RT formula in
 the surface/state correspondence \cite{Miyaji2015a}.} 

It is illuminating to compare the cut-off surfaces with the layers in cMERA. The coarse graining in cMERA is reflected in  the quantum states
in different layers. On the other hand, the cut-off surfaces corresponding to different   $T\overline{T}$ deformations play similar role, as shown in the holographic RG flow argument\cite{McGough2016}.

\section{Surface/State correspondence, $T\overline{T}$ deformation and complexity}

Complexity captures the information beyond entanglement.
In this section, we study the complexity of the ground state of $T\overline{T}$
deformed CFT and compare it with the \textcolor{black}{complexity of the} cMERA which has been
studied in \cite{Chapman2017,Molina2018}. \textcolor{black}{The holographic
complexity of the ground state is studied as well.}

We follow the definition of complexity for field theory from \cite{Chapman2017,Molina2018}.
 For two quantum states $\ensuremath{|\Phi(\xi)\rangle}$
and $\ensuremath{|\Phi(\xi+d\xi)\rangle}$ where $\ensuremath{\xi=(\xi_{1},\xi_{2},\dots,\xi_{n})}$
\textcolor{black}{are} the parameters of \textcolor{black}{the} state space, the (squared) Hilbert-Schmidt distance
is defined by
\begin{equation}
d_{HS}^{2}(\xi)\equiv1-|\langle\Phi(\xi+d\xi)|\Phi(\xi)\rangle|^{2}=g_{ij}d\xi_{i}d\xi_{j},\label{hsdistance}
\end{equation}
in which $g_{ij}$ is the Fisher information metric. For two points
$\ensuremath{|\Phi(0)\rangle}$ and $\ensuremath{|\Phi(\lambda_{max})\rangle}$
in the state space, one can find a curve $\ensuremath{|\Phi(\lambda)\rangle}$
connecting them which has the minimum length \textcolor{black}{under} the Fisher information
metric. 
Supposing the reference state
$|\Psi\rangle=\ensuremath{|\Phi(0)\rangle}$ and the target state
$|\Upsilon\rangle=\ensuremath{|\Phi(\lambda_{max})\rangle}$, then
the complexity between $|\Upsilon\rangle$ and $|\Psi\rangle$ is
defined by
\begin{equation}
C_{(|\Psi\rangle,|\Upsilon\rangle)}=\int_{0}^{\lambda_{max}}d_{HS}(\lambda)=\int_{0}^{\lambda_{max}}\sqrt{g_{\lambda\lambda}}d\lambda.\label{eq:Cgamma}
\end{equation}
Note that (\ref{eq:Cgamma}) is invariant under the reparametrization
of $\lambda$.

\subsection{The complexity of cMERA}

The definition (\ref{eq:Cgamma}) seems \textcolor{black}{good} but difficult to apply
due to the very high dimensions of the state space.
Fortunately, the situation is
simplified for cMERA. The layers in cMERA can be labeled by one parameter
$u$ and the corresponding quantum states by $\ensuremath{|\Phi(u)\rangle}$.
We suppose that the curve composed by $\ensuremath{|\Phi(u)\rangle}$
is exactly the minimum curve for computing the complexity and the
metric along this curve is just the Fisher information metric, even though
 more choices exist. According to \cite{Molina2018}, one can write
the line element as
\begin{equation}
d_{HS}^{2}(u)\equiv1-|\langle\Phi(u+du)|\Phi(u)\rangle|^{2}=\mathcal{N}(u)g_{uu}du^{2}.
\end{equation}
 Here
 \be
 \mathcal{N}(u)=\textrm{Vol}\cdot\int_{|k|\leq\Lambda e^{u}}d^{d}k, \ee
 is the volume of the effective phase space at $u$-th layer,
in which $d$ is the spacial dimension, $\textrm{Vol}$ is the volume of $R^{d}$,
$k$ is the momentum, $\Lambda=1/\epsilon$ is the ultraviolet cutoff
and $\Lambda e^{u}$ is the momentum cutoff for the $u$-th layer.
Using $\mathcal{N}(u)$, one can define a series of  complexity of cMERA
 \cite{Chapman2017,Molina2018} by
\begin{align}
C^{(n)} & =\int\mathcal{N}(u)^{\frac{1}{n}}\sqrt{g_{uu}}du.
\end{align}
Note that $C^{(2)}$ is just the complexity defined by Fisher information
metric. For the ground state of 1+1 dimensional massless free scalar,
\textcolor{black}{its cMERA complexity} has been worked out as
\begin{equation}
C_{\textrm{cMERA}}^{(1)}=\frac{\textrm{Vol}\cdot\Lambda}{2},\ \ C_{\textrm{cMERA}}^{(2)}=\sqrt{\textrm{Vol}\cdot\Lambda}, \label{eq:cMERAn}
\end{equation}
 where Vol is the length of $R^{1}$ and $\Lambda$ is the momentum
cutoff of cMERA. $\Lambda$ can also be understood as the momentum
cutoff at different layer $u$ in the same cMERA. So we
have
\begin{equation}
C_{|\Phi(u)\rangle}^{(1)}=\frac{\textrm{Vol}\cdot\Lambda e^{u}}{2},\ \ C_{|\Phi(u)\rangle}^{(2)}=\sqrt{\textrm{Vol}\cdot\Lambda e^{u}},
\end{equation}
 where $\Lambda e^{u}$ is the momentum cutoff at $u$-th layer.

\subsection{The complexity of the ground state of $T\overline{T}$ deformed CFT}

In the case of  the $T\overline{T}$-deformed CFT, the Hilbert-Schmidt distance between the ground states of theories $\mathcal{T}^{(\mu)}$
 and $\mathcal{T}^{(\mu+d\mu)}$ is
\begin{equation}
d_{HS}^{2}(\mu)\equiv1-|\langle\Omega^{(\mu+d\mu)}|\Omega^{(\mu)}\rangle|^{2}=\mathcal{N}(\mu)g_{\mu\mu}d\mu^{2},\label{hsdistance2}
\end{equation}
 where $\mathcal{N}(\mu)$ is similar to $\mathcal{N}(u)$,
 \be
  \mathcal{N}(\mu)=\textrm{Vol}/\sqrt{\mu} \ee
with $1/\sqrt{\mu}$ being the momentum cutoff of theory $\ensuremath{\mathcal{T}^{(\mu)}}$
and $\textrm{Vol}$ being the length of $R^{1}$. Then the complexity of the ground
state of $T\overline{T}$-deformed 2D CFT could be defined as
\begin{align}
C^{(n)}(\mu) & =\int\mathcal{N}(\mu)^{\frac{1}{n}}\sqrt{g_{\mu\mu}}d\mu. \label{ttcomplexity}
\end{align}
 Once the metric $g_{\mu\mu}$ is obtained from (\ref{hsdistance2}),
one can get the complexity of $\ensuremath{|\Omega^{(\mu)}\rangle}$
from above formula.

The distance (\ref{hsdistance2})
can be calculated by using the method \textcolor{black}{in} \cite{Miyaji2015c}. We
consider the $T\overline{T}$-deformed CFT living on the plane. Then
$|\langle\Omega^{(\mu+d\mu)}|\Omega^{(\mu)}\rangle|^{2}$ can be expressed
by the path integral
\begin{align}
&\langle\Omega^{(\mu+d\mu)}|\Omega^{(\mu)}\rangle \nonumber\\
=&\frac{\int D\phi\exp[-\int dx(\int_{-\infty}^{0}dt\mathcal{L}^{(\mu)}+\int_{0}^{\infty}dt\mathcal{L}^{(\mu+d\mu)})]}{\sqrt{Z^{(\mu)}Z^{(\mu+d\mu)}}}
\end{align}
 where $Z^{(\mu)}$ is the partition function
 of theory $\ensuremath{\mathcal{T}^{(\mu)}}$. According to the definition of $T\overline{T}$ deformation, we have
\begin{align}
\mathcal{L}^{(\mu+d\mu)}-\mathcal{L}^{(\mu)} & \equiv\delta\mathcal{L}=d\mu\cdot O^{(\mu)}(t,x).
\end{align}

Introducing an UV cutoff $\epsilon$, the ground state becomes
\begin{equation}
|\tilde{\Omega}^{(\mu+d\mu)}(\epsilon)\rangle\equiv\frac{e^{-\epsilon H^{(\mu)}}|\Omega^{(\mu+d\mu)}\rangle}{(\langle\Omega^{(\mu+d\mu)}|e^{-2\epsilon H^{(\mu)}}|\Omega^{(\mu+d\mu)}\rangle)^{1/2}},
\end{equation}
 where $H^{(\mu)}$ is the Hamiltonian of the theory $\ensuremath{\mathcal{T}^{(\mu)}}$. The cutoff $\epsilon$ depends on $\mu$
implicitly  $\epsilon\sim\sqrt{\mu}$. Then we get
\begin{equation}
\langle\tilde{\Omega}^{(\mu+d\mu)}(\epsilon)|\Omega^{(\mu)}\rangle=\frac{\langle\exp[-\int_{\epsilon}^{\infty}dt\int dx\delta\mathcal{L}]\rangle}{(\langle\exp[-(\int_{-\infty}^{-\epsilon}+\int_{\epsilon}^{\infty})dt\int dx\delta\mathcal{L}]\rangle)^{1/2}},
\end{equation}
 where $\ensuremath{\langle\dots\rangle}$ denotes the expectation value in the
ground state $\ensuremath{|\Omega^{(\mu)}\rangle}$. Expanding the
above formula to the second order of $d\mu$, one gets
\begin{align}
& 1-|\langle\tilde{\Omega}^{(\mu+d\mu)}(\epsilon)|\Omega^{(\mu)}\rangle|^{2}  \label{eq:dCFTds}\\
=&d\mu^{2}\int_{\epsilon}^{\infty}dt\int_{-\infty}^{-\epsilon}dt^{\prime}\int dx dx^{\prime}\langle O^{(\mu)}(t,x)O^{(\mu)}(t^{\prime},x^{\prime})\rangle.\nonumber
\end{align}
 Here the time reversal symmetry $\langle\delta\mathcal{L}(-t,x)\delta\mathcal{L}(-t^{\prime},x^{\prime})\rangle=\langle\delta\mathcal{L}(t,x)\delta\mathcal{L}(t^{\prime},x^{\prime})\rangle$
was used.

When $\mu=0$, one has $O^{(0)}=T\overline{T}$ where $T,\overline{T}$
are the energy-momentum tensor of the original CFT. So we have
\begin{align}
\langle O^{(0)}(t,x)O^{(0)}(t^{\prime},x^{\prime})\rangle  
  =\frac{c^{2}/4}{((t-t^{\prime})^{2}+(x-x^{\prime})^{2})^{4}}.\label{eq:OO0}
\end{align}
The general expression of $\langle O^{(\mu)}(t,x)O^{(\mu)}(t^{\prime},x^{\prime})\rangle$ is unknown. However,
its behavior can be obtained by \textcolor{black}{dimensional} analysis. {Since $(\mu,t,x)$ is equivilent
to $(\lambda^{2}\mu,\lambda t,\lambda x)$, we find that the correlation function takes the form}
\begin{equation}
\langle O^{(\mu)}(t,x)O^{(\mu)}(t^{\prime},x^{\prime})\rangle=
\frac{c^{2}[1+\sum_{i=1}^{\infty}a_{i}
(\frac{\mu}{(t-t^{\prime})^{2}+(x-x^{\prime})^{2}})^{i}]}
{4((t-t^{\prime})^{2}+
(x-x^{\prime})^{2})^{4}}
,\label{eq:OOmu1}
\end{equation}
 where the coefficients $a_{i}$ are constants that cannot be determined \textcolor{black}{now}. 

Plugging (\ref{eq:OOmu1}) into (\ref{eq:dCFTds}), we get
\begin{equation}
1-|\langle\tilde{\Omega}^{(\mu+d\mu)}(\epsilon)|\Omega^{(\mu)}\rangle|^{2}=\mathcal{N}(\mu)g_{\mu\mu}d\mu^{2}=N\cdot\textrm{Vol}\cdot\mu^{-5/2}d\mu^{2}.
\end{equation}
 Here $\epsilon\sim\sqrt{\mu}$ was used. $N$ is a dimensionless constant
and $\textrm{Vol}$ is the length of $R^{1}$. Since $\ensuremath{\mathcal{N}(\mu)=\textrm{Vol}/\sqrt{\mu}}$,
we get
\begin{equation}
g_{\mu\mu}=\frac{N}{\mu^{2}}.
\end{equation}
 Plugging these into the definition of complexity (\ref{ttcomplexity}),
we get
\begin{align}
C^{(n)}(\mu) & =\int\mathcal{N}(\mu)^{\frac{1}{n}}\sqrt{g_{\mu\mu}}d\mu\propto\bigg(\frac{\textrm{Vol}}
{\sqrt{\mu}}\bigg)^{\frac{1}{n}}.\label{ttcomplexity1}
\end{align}
 Now we compare the above result with the complexity of cMERA (\ref{eq:cMERAn}).
 Since $1/\sqrt{\mu}$
is the momentum cutoff of theory $\ensuremath{\mathcal{T}^{(\mu)}}$,
the results are consistent given $\Lambda\leftrightarrow1/\sqrt{\mu}$.
Different ground state $\ensuremath{|\Omega^{(\mu)}\rangle}$ in $T\overline{T}$
deformed CFT corresponds to $\ensuremath{|\Phi(u)\rangle}$ at different
layer $u$ in cMERA. This supports our proposal that cAdS/dCFT is
a realization of surface/state correspondence. More concretely, the
corresponding quantum state of surface $\ensuremath{\mathcal{S}^{r_{c}}}$
is the ground state $\ensuremath{|\Omega^{(\mu)}\rangle}$ in theory
$\ensuremath{\mathcal{T}^{(\mu)}}$.

\subsection{Holographic complexity}

\begin{figure}[htbp!]
\centering{}\includegraphics[width=0.25\textwidth]{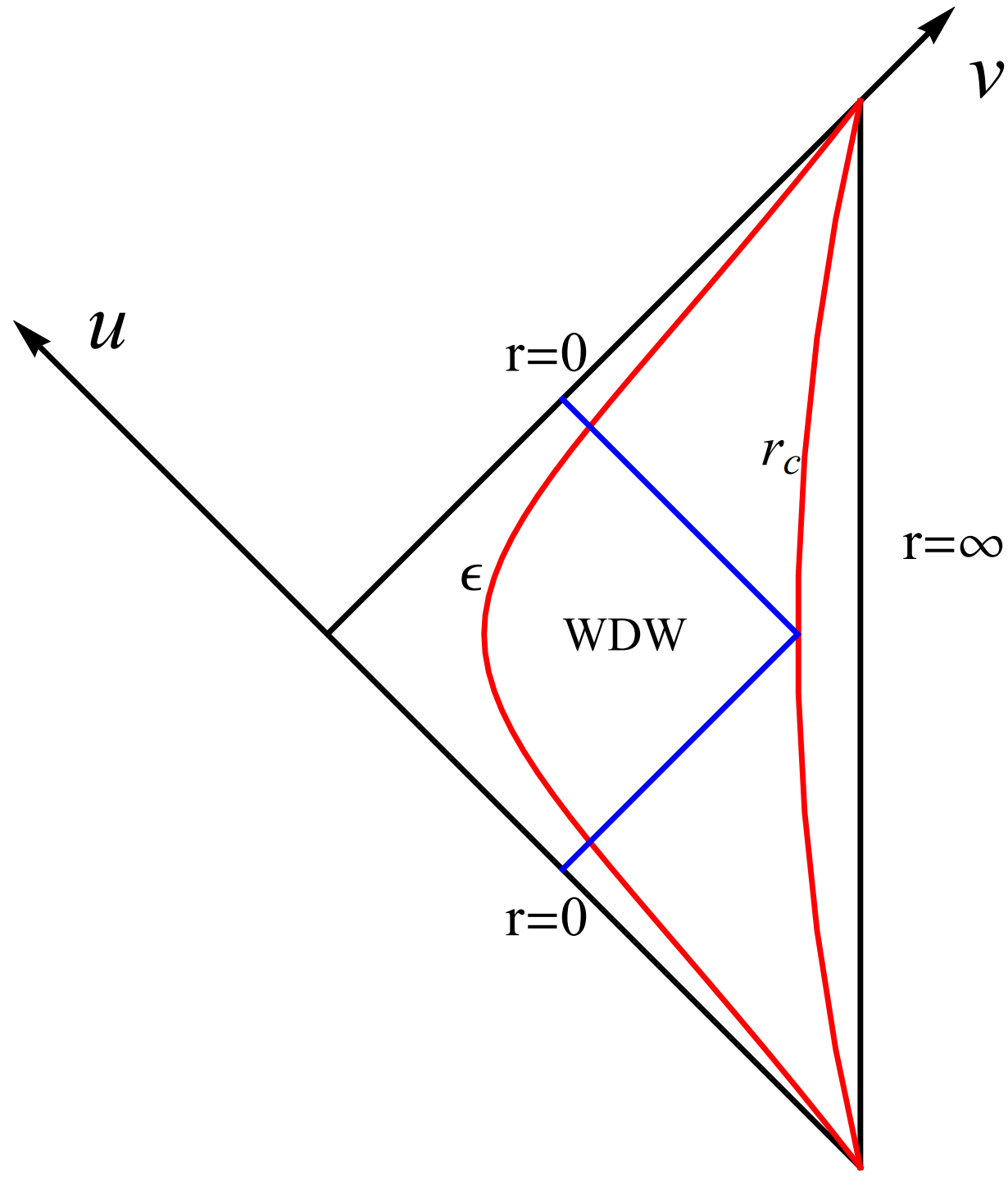}\caption{\label{WDW} The WdW patch for the Poincare patch of $AdS_{3}$. We
introduced a regularized surface $r=\epsilon$.}
\end{figure}

There are two conjectures about the holographic complexity: CV conjecture
and CA conjecture. The CV conjecture claims that the complexity of
a quantum state \textcolor{black}{on the boundary time slice $\Xi$} is proportional to the maximum volume of the spacelike
hypersurface \textcolor{black}{which meets the boundary on $\Xi$}:
\begin{equation}
C_{V}(\Xi)=\max_{\Xi=\partial B}\bigg[\frac{\mathcal{V}(B)}{G_{N}L}\bigg].\label{CV}
\end{equation}
 Here $B$ is \textcolor{black}{a hypersurface which meets the boundary on $\Xi$} and $\mathcal{V}(B)$
is its volume. The CA conjecture claims that the complexity of a quantum
state \textcolor{black}{on the boundary time slice $\Xi$} is
\begin{equation}
C_{A}(\Xi)=\frac{I_{\textrm{WdW}}}{\pi\hbar}.
\end{equation}
Here $I_{\textrm{WdW}}$ is the onshell action in the WdW patch which
is the closure of all spacelike surfaces with boundary $\Xi$. The
WdW patch for the Poincare patch of $AdS_{3}$ we considered here
is shown in Figure \ref{WDW}.

\subsubsection{The CV conjecture}

The $T\overline{T}$ deformed CFT living on a 2-dimensional plane
is dual to the Poincare patch of $AdS_{3}$ with boundary at finite
cutoff:
\begin{equation}
ds^{2}=-\frac{r^{2}}{L^{2}}dt^{2}+\frac{L^{2}}{r^{2}}dr^{2}+\frac{r^{2}}{L^{2}}dx^{2},\;\;(r\leq r_{c}).\label{adspatch}
\end{equation}
 The deformed theory $\ensuremath{\mathcal{T}^{(\mu)}}$
lives on the new boundary $r=r_{c}$ satisfying (\ref{muandrc}).
Its Penrose diagram is shown in Figure \ref{WDW}.

We are only interested in the complexity of the ground state $\ensuremath{|\Omega^{(\mu)}\rangle}$
in the theory $\ensuremath{\mathcal{T}^{(\mu)}}$. Since $\ensuremath{|\Omega^{(\mu)}\rangle}$
is just the quantum state at time $t=0$ in theory $\ensuremath{\mathcal{T}^{(\mu)}}$,
the boundary $\Xi$ is at $(t=0,r=r_{c})$. \textcolor{black}{Due to the} symmetry, $B$
is the hypersurface with $t=0$. Its volume is
\begin{equation}
\mathcal{V}(B)=\int dx\int_{0}^{r_{c}}dr\sqrt{\frac{L^{2}}{r^{2}}\frac{r^{2}}{L^{2}}}=\textrm{Vol}\cdot r_{c},
\end{equation}
 where $\textrm{Vol}$ is the length of $R^{1}$. So the complexity
is
\begin{equation}
C_{V}(|\Omega^{(\mu)}\rangle)=\frac{\textrm{Vol}\cdot r_{c}}{G_{N}L}\propto\frac{\textrm{Vol}}{\sqrt{\mu}},
\end{equation}
 where the relation (\ref{muandrc}) was used. Comparing  with the result (\ref{ttcomplexity1})
from field theory, we find that
\begin{equation}
C_{V}(|\Omega^{(\mu)}\rangle)\sim C^{(1)}(\mu).
\end{equation}
 This implies that $C^{(1)}$ is  \textcolor{black}{a more} suitable definition of complexity
for field theory.

\subsubsection{The CA conjecture}

To \textcolor{black}{investigate} the CA conjecture, it is convenient to introduce the tortoise coordinate
$
r^{\ast}(r)=\int\frac{L^{2}}{r^{2}}dr=\frac{-L^{2}}{r}.
$
 We construct the Eddington-Finkelstein outgoing and in-falling coordinates
as
$
u =t-r^{\ast}(r),\
v =t+r^{\ast}(r)
$
respectively.  Then the metric of the Poincare patch of $AdS_{3}$ can be written
as
\begin{align}
ds^{2} & =-f(r)du^{2}-2dudr+\frac{r^{2}}{L^{2}}dx^{2}\\
 & =-f(r)dv^{2}+2dvdr+\frac{r^{2}}{L^{2}}dx^{2},
\end{align}
where $f(r)=r^{2}/L^{2}$. The past null boundary of the WdW patch
for time slice $(t=0,r=r_{c})$  is
$
u=u_{c}=\frac{L^{2}}{r_{c}}.
$
 The future null boundary is
$
v=v_{c}=-\frac{L^{2}}{r_{c}}.
$


The calculation of CA conjecture is straightforward and the final result is
\begin{equation}
C_{A}(|\Omega^{(\mu)}\rangle)=\frac{\textrm{Vol}\cdot r_{c}}{8\pi^{2}\hbar G_{N}L}\log\frac{\tilde{L}^{2}}{L^{2}}.
\end{equation}
Here $\tilde{L}$ is a constant introduced to \textcolor{black}{remove the ambiguity associated with the} normalization of the tangent
vectors of the null boundaries \cite{Lehner2016}.
 Comparing with the result from
CV conjecture, we find that they are consistent up to a constant factor
$\frac{1}{8\pi}\log\frac{\tilde{L}^{2}}{L^{2}}$. We should require
$\tilde{L}>L$ to ensure they have the similar behavior. Comparing
the result of CA conjecture with that from field theory, we find
\begin{equation}
C_{A}(|\Omega^{(\mu)}\rangle)\sim C^{(1)}(\mu).
\end{equation}
 Thus for the ground state of $T\overline{T}$ deformed CFT, the complexity
defined by $C^{(1)}$ is consistent with the corresponding holographic
CA and CV conjectures.

\section{Conclusions and discussions}
{The surface/state correspondence is an interesting proposal to study the emergence of spacetime.
 In this letter, we propose that cAdS/dCFT
correspondence is a realization of surface/state correspondence. More
concretely, when the surface is a time slice \textcolor{black}{of the moved boundary},
the corresponding state  is just
the ground state of the $T\overline{T}$-deformed CFT. This proposal  is novel.
\textcolor{black}{The study of the entanglement entropy of the $T\overline{T}$-deformed CFT gives us the first evidence for our proposal.} We further tested our proposal from the {perspective} of the  complexity. The complexity
of the ground state of $T\overline{T}$-deformed CFT is consistent
with that of the cMERA \textcolor{black}{for the ground state of a massless free scalar}, and also with the holographic complexity.} 

In this work, we only studied the complexity of the ground state \textcolor{black}{of the $T\overline{T}$ deformed CFT}. {To test our proposal about defining the complexity via $T\overline{T}$ deformation,
it would be interesting to generalize our study to thermofield double states in the future. Since the thermofield double state is dual to the eternal BTZ black hole, we can check whether the complexity of the thermofield double state obtained by our method is consistent with the holographic complexity of the eternal BTZ black hole.}



\section*{Acknowledgements}

The work was in part supported by NSFC Grant No. 11275010, No. 11335012,
No. 11325522 and No. 11735001.

\onecolumngrid

\twocolumngrid

\end{document}